\documentclass[12pt]{article}
\usepackage{amsmath,amssymb}
\usepackage{graphicx}
\usepackage[utf8]{inputenc}
\usepackage{booktabs}
\usepackage{textcomp}
\usepackage{authblk}
\usepackage{stfloats}
\usepackage[section]{placeins}
\usepackage{setspace}
\usepackage[margin=2.5cm]{geometry}
\usepackage[super,comma,sort&compress]{natbib}
\usepackage[breaklinks]{hyperref}
\hypersetup{colorlinks=true,linkcolor=black,citecolor=black,urlcolor=blue}
\begin{document}

\title{Foundation-model-guided radiogenomic discovery linking cancer genomes to cancer scans}
\author[1,*]{Frederik Hauke}
\author[2]{Jeremias Krause}
\author[1]{Patrick Wienholt}
\author[1]{Christiane Kuhl}
\author[2]{Ingo Kurth} 
\author[6]{Sikander Hayat}
\author[3,4,5]{Jakob Nikolas Kather}
\author[1]{Sven Nebelung}
\author[1]{Daniel Truhn}

\affil[1]{Department of Diagnostic and Interventional Radiology, University Hospital RWTH Aachen, Aachen, Germany}
\affil[2]{Center for Human Genetics and Genomic Medicine, University Hospital RWTH Aachen, Aachen, Germany}
\affil[3]{Else Kr\"oner Fresenius Center for Digital Health, TU Dresden, Dresden, Germany}
\affil[4]{Department of Medicine I, Faculty of Medicine and University Hospital Carl Gustav Carus, TUD Dresden University of Technology, Dresden, Germany}
\affil[5]{Department of Medical Oncology, National Center for Tumor Diseases (NCT), Heidelberg University Hospital, Heidelberg, Germany}
\affil[6]{Department of Nephrology, Rheumatology and Immunology (Medical Clinic II), University Hospital RWTH Aachen, Aachen, Germany}
\affil[*]{Corresponding author: fhauke@ukaachen.de}

\begin{abstract}
The function of many genes is still unknown, and conventional driver-discovery methods, which rely on how frequently a gene is mutated, cannot assess genes that are only rarely affected. Here we pair Evo~2-based genome analysis with routine clinical imaging to identify gene--phenotype associations at genome-wide scale. For every somatic mutation across three TCGA cohorts (cRCC=clear cell renal cell carcinoma, HCC=hepatocellular carcinoma, and BC=breast cancer; $n = 340$ total), Evo~2 predicts a severity score, with no task-specific training. Per-gene severity summaries are then correlated with radiomic features extracted from paired tumor segmentations, controlling for total mutation burden. In TCGA-cRCC ($n = 162$), this sweep recovers established renal-cancer drivers and identifies 46 additional genes reaching false discovery rate (FDR) significance absent from curated cancer-gene panels, several of which are Mendelian ciliopathy and cytoskeletal-disease genes. These results demonstrate that pairing a genomic language model with widely available clinical imaging can serve as a hypothesis-free discovery tool for gene--imaging associations invisible to conventional approaches.
\end{abstract}

\maketitle

\section{Introduction}

Cancer arises from somatic mutations that co-opt pathways governing growth, survival, and tissue organisation \cite{pan_cancer_whole_genomes_2020,nrg3539}. Large-scale sequencing has mapped this mutational landscape in detail: the Pan-Cancer Analysis of Whole Genomes (PCAWG) consortium reported 4--5 driver mutations per genome across 2{,}658 tumors \cite{pan_cancer_whole_genomes_2020}, exome surveys catalogued drivers in 727 cancer genes across more than 20{,}000 primary tumors \cite{mutational_landscape_drivers_2023}, and whole-genome sequencing of 10{,}478 patients identified 330 candidate driver genes, 74 previously unlinked to any cancer \cite{dietlein_10478_cancer_genomes_natgen_2024}. Despite this wealth of catalogued variants, the phenotypic consequences of most somatically mutated genes remain poorly understood. Current driver-discovery methods rely on recurrence statistics that demand large cohorts and offer limited insight into variant-level effects \cite{somatic_mutation_rates_natbiotech_2022}, while thousands of protein-coding genes still lack reliable functional annotation altogether \cite{unknome_plos_biology_2023}. Linking these genes to phenotypes will require scalable approaches that look beyond mutation recurrence alone.

Medical imaging provides a complementary window into tumor biology. Radiomics, the high-throughput extraction of quantitative features from clinical images, captures variation in intensity, shape, and texture that reflects underlying tissue architecture and gene-expression programmes \cite{radiomics_tumor_phenotype_natcomm_2014}. Because such images are acquired routinely for virtually every cancer patient, they represent a phenotypic data source collected at scale and at low cost that is largely untapped for genomic discovery.

Several studies have begun to bridge the gap between imaging phenotypes and tumor genotype. An integrated analysis of 763 lung adenocarcinomas demonstrated that somatic mutations in \textit{EGFR} and \textit{KRAS} drive distinct computed tomography (CT)-derived radiomic phenotypes, with a signature distinguishing \textit{EGFR}-positive from \textit{KRAS}-positive tumors at an area under the receiver operating characteristic curve (AUC) of 0.80 \cite{0008_5472_can_17_0122v2}. A genome-wide association approach applied to pancreatic cancer linked sets of driver genes to specific radiomic features, establishing that somatic mutations shape tumor morphological heterogeneity in a manner detectable by imaging \cite{radiogwas_pancreatic_2024}. These efforts, however, remain confined to known driver genes identified through recurrence statistics; they do not address the vast majority of somatically mutated genes whose functional consequences are unknown.

Meanwhile, genomic foundation models have emerged that can predict the functional impact of genetic variants directly from sequence, without task-specific training. DNA language models, trained on large collections of genomic sequences, learn evolutionary constraints and sequence context that allow zero-shot prediction of variant effects \cite{nucleotide_transformer_2024,alphagenome_nature_2025}. Evo~2 assigns a probability to any genomic sequence at single-nucleotide resolution; it is a biological foundation model trained on 9.3 trillion DNA base pairs from a curated atlas spanning all domains of life \cite{s41586_026_10176_5_3}. By comparing the model's log-likelihood for a reference and a mutant allele, one obtains a zero-shot severity score that distinguishes pathogenic from benign ClinVar variants in both coding and non-coding regions and separates loss-of-function from functional alleles in \textit{BRCA1} saturation mutagenesis data \cite{s41586_026_10176_5_3}. A recent population-genomics study showed that Evo~2 scores can prioritise putatively functional variants and quantify haplotype effects on Alzheimer's disease endophenotypes across diverse ancestry cohorts \cite{v1_covered_88832dba_a70d_41a6_b6bc_435eb56fee87}. These capabilities make genomic foundation models attractive for scoring somatic mutations at scale, including in genes for which no prior functional annotation exists.

\begin{figure*}[!htb]
\centering
\includegraphics[width=\textwidth]{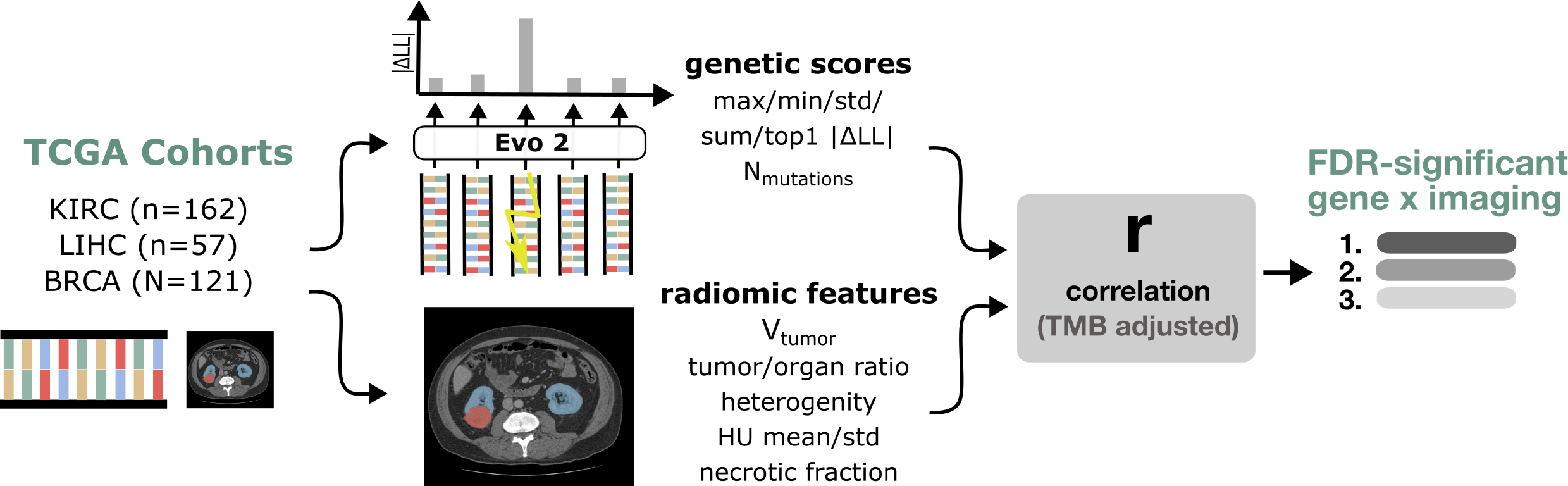}
\caption{Overview of the cross-modal analysis pipeline. Somatic mutations from three TCGA cohorts (cRCC, $n=162$; HCC, $n=57$; BC, $n=121$) are scored using the Evo~2 genomic language model, which assigns a log-likelihood drop $|\Delta\mathrm{LL}|$ to each variant. Per-gene severity summaries (maximum, mean, sum, standard deviation, signed extremes, and mutation count) are computed for every gene with at least five carriers. In parallel, radiomic features (tumor volume, tumor-to-organ ratio, intensity heterogeneity, intensity mean and standard deviation, and necrotic fraction) are extracted from paired tumor segmentations. A TMB-adjusted partial Spearman correlation is then computed for each combination of per-gene severity metric and imaging feature to identify FDR-significant gene--imaging associations.}
\label{fig:workflow}
\end{figure*}

Here we ask whether a genomic foundation model and routine clinical imaging can be combined into a single discovery framework for uncovering gene--phenotype associations at genome-wide scale (Fig.~\ref{fig:workflow}). We use Evo~2 to score every somatic mutation in three TCGA cohorts (cRCC=clear cell renal cell carcinoma, HCC=hepatocellular carcinoma, and BC=breast cancer), aggregate per-gene severity summaries, and correlate them with radiomic features extracted from paired tumor segmentations, controlling for total mutation burden (TMB). In the largest cohort (TCGA-cRCC, $n=162$), this cross-modal sweep recovers established renal cancer drivers at genome-wide significance and, more importantly, identifies 46 genes outside any curated cancer-gene panel whose Evo~2 severity profiles correlate with tumor volume. Several of these hits are well-characterised Mendelian disease genes involved in ciliopathy and cytoskeletal biology, positioning them as candidate links between inherited structural-biology programmes and somatic tumor phenotype. Our results demonstrate that combining a general-purpose genomic language model with ubiquitous imaging data can serve as a hypothesis-free discovery tool, revealing gene--imaging associations that recurrence-based driver analyses would miss.

\section{Results}

\subsection{Cohort and analysis overview}

After paired-modality filtering, the three cohorts contributed 162 (cRCC), 57 (HCC) and 121 (BC) patients with both somatic mutation calls and tumor imaging. The TMB-adjusted partial-Spearman sweep yielded 285 FDR-significant per-gene metric--imaging feature combinations from 65 unique genes meeting $\text{FDR} < 0.05$ in cRCC; under the same family-wise control neither HCC nor BC produced a single FDR-significant hit, reflecting their reduced sample size.

\begin{figure*}[!htb]
\centering
\includegraphics[width=\textwidth]{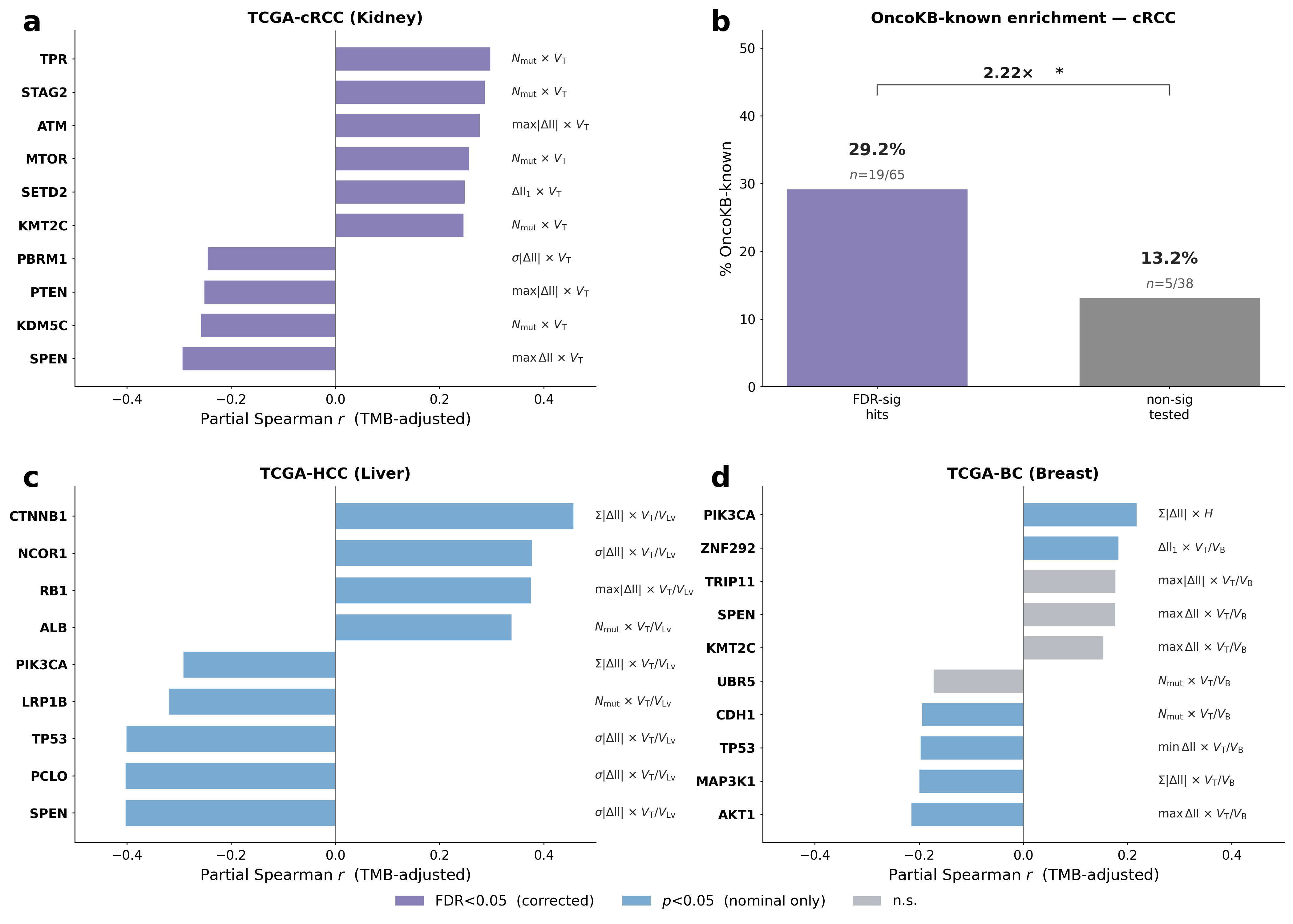}
\caption{Rediscovery of established cancer drivers through Evo~2--imaging correlation. \textbf{(a)}~Top-10 OncoKB-known genes in TCGA-cRCC ranked by FDR-corrected significance, showing TMB-adjusted partial Spearman $r$ between each gene's Evo~2 severity metric and the corresponding radiomic feature (annotated at right). All bars reach $\text{FDR} < 0.05$ (purple). \textbf{(b)}~Enrichment of OncoKB-known genes among FDR-significant cRCC hits (29.2\%, 19/65) versus non-significant tested genes (13.2\%, 5/38), yielding a 2.22-fold enrichment (Fisher's exact $p = 0.05$). \textbf{(c)}~Top-10 OncoKB-known genes in TCGA-HCC by uncorrected $p$; none reaches cohort-level FDR significance (blue, $p < 0.05$ nominal only). \textbf{(d)}~Top-10 OncoKB-known genes in TCGA-BC by uncorrected $p$; none reaches FDR significance. Grey bars indicate $p \geq 0.05$.}
\label{fig:rediscovery}
\end{figure*}

\subsection{Rediscovery of OncoKB-known cancer drivers}

Of the 65 cRCC hits, 19 fall within the OncoKB-known panel; the remaining 46 are outside any curated cancer-gene set. The top-ranked OncoKB-known cRCC genes (Fig.~\ref{fig:rediscovery}a) span the established renal-cancer landscape: chromatin regulators (\textit{PBRM1}, \textit{SETD2}, \textit{KDM5C}, \textit{KMT2C}) and the cohesin subunit \textit{STAG2} \cite{nrg3539,tcga_kirc_nature_2013,dietlein_10478_cancer_genomes_natgen_2024}, PI3K-pathway and DNA-damage-response components (\textit{PTEN}, \textit{MTOR}, \textit{ATM}) \cite{tcga_kirc_nature_2013}, and the nucleoporin \textit{TPR}. Each of these reaches FDR-significant correlation, predominantly with tumor volume.

HCC and BC produced no OncoKB-known gene at $\text{FDR} < 0.05$, but the top-10 OncoKB-known genes ranked by uncorrected $p$ (Fig.~\ref{fig:rediscovery}c,~d) recover the canonical drivers of each disease (\textit{CTNNB1}, \textit{TP53}, \textit{RB1} and \textit{PIK3CA} in the liver cohort; \textit{PIK3CA}, \textit{CDH1}, \textit{TP53}, \textit{KMT2C} and \textit{AKT1} in the breast cohort \cite{dietlein_10478_cancer_genomes_natgen_2024,mutational_landscape_drivers_2023}), even though none survives the strict per-cohort FDR threshold.

To verify that FDR separates a biologically meaningful subset of genes from the larger tested pool rather than splitting it at random, we tested whether the FDR-significant cRCC hits are enriched for OncoKB-known cancer drivers relative to the non-significant tested genes (Fig.~\ref{fig:rediscovery}b). Both arms passed identical filtering and the same severity-metric-by-imaging-feature correlation machinery; they differ only in whether they crossed FDR. The FDR-significant set was 2.22-fold enriched for OncoKB-known genes (29.2\% vs 13.2\%; Fisher's exact, one-sided, $p = 0.05$), confirming that the threshold isolates a panel-enriched subpopulation. Together, these results demonstrate that combining a genomic foundation model with routine clinical imaging recovers the known driver landscape of renal, liver and breast cancer without any cancer-specific supervision, validating the framework before turning to genes outside curated panels.

\begin{figure*}[!htb]
\centering
\includegraphics[width=\textwidth]{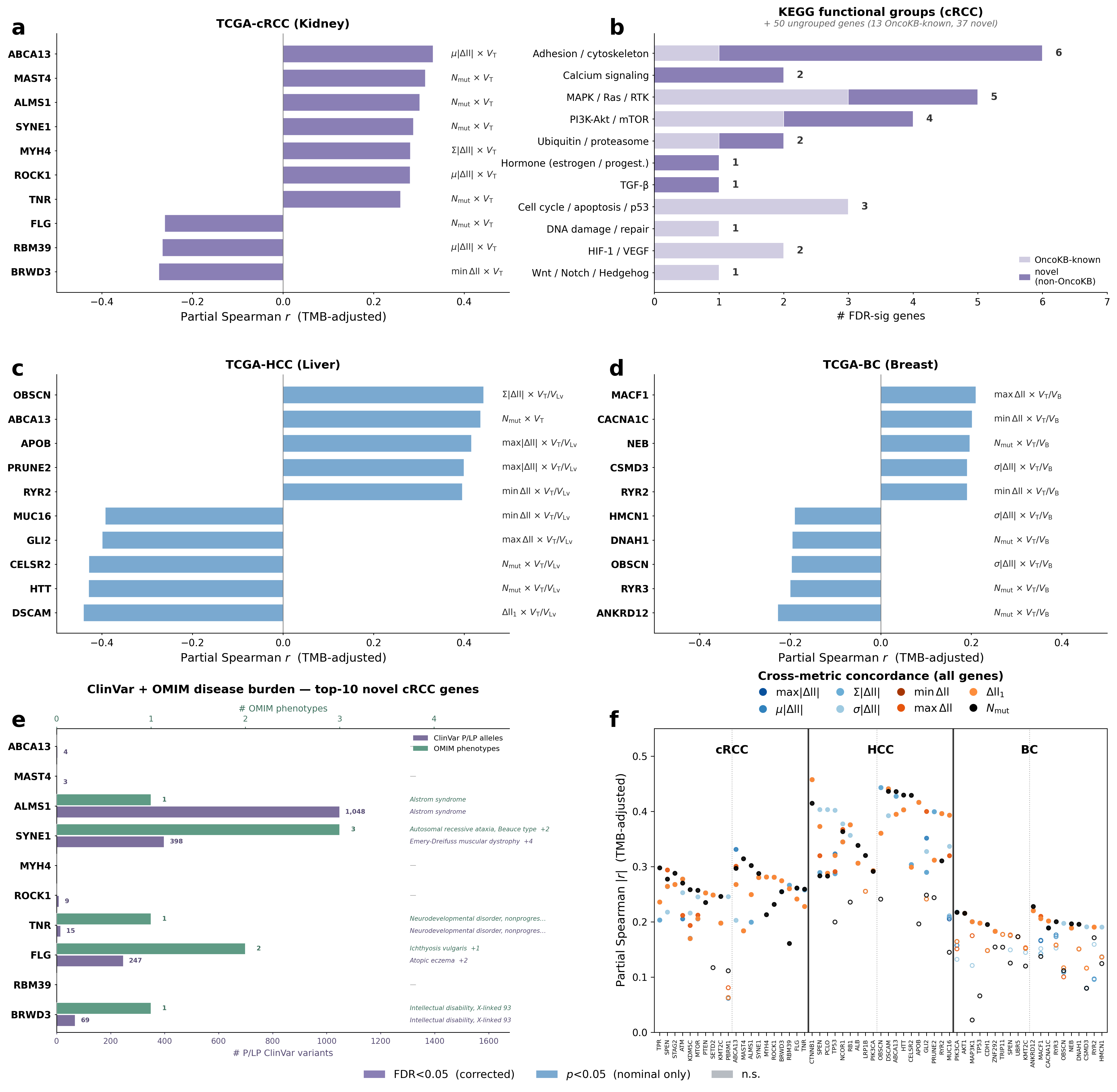}
\caption{Novel gene discoveries, pathway context and robustness. \textbf{(a)}~Top-10 non-OncoKB cRCC genes at $\text{FDR} < 0.05$, ranked by significance; bars show TMB-adjusted partial Spearman $r$ with the radiomic feature noted at right. \textbf{(b)}~The 65 FDR-significant cRCC genes mapped to eleven KEGG functional groups, split by OncoKB-known (light purple) and novel (dark purple); genes in multiple groups are counted in each. \textbf{(c,\,d)}~Top-10 non-OncoKB genes in HCC and BC by uncorrected $p$; none reaches FDR significance. \textbf{(e)}~ClinVar pathogenic/likely-pathogenic allele counts (lavender, bottom axis) and OMIM phenotype counts (teal, top axis) for the ten genes in~(a); five are established Mendelian-disease genes, five have no entry in either database. \textbf{(f)}~Cross-metric concordance: per-gene $|r|$ with the headline radiomic feature for all eight Evo~2 severity metrics (filled = $p < 0.05$). The four $|\Delta\mathrm{LL}|$ magnitude summaries are near-identical within each gene; the mutation count $N_{\mathrm{mut}}$ diverges most.}
\label{fig:discovery}
\end{figure*}

\subsection{Discovery of novel imaging-correlating genes}

The 46 cRCC discoveries outside OncoKB show robust correlations with tumor-volume features (Fig.~\ref{fig:discovery}a). The strongest novel hits include the ABC-family transporter \textit{ABCA13}, the microtubule-associated serine/threonine kinase \textit{MAST4}, the ciliopathy gene \textit{ALMS1}, the nuclear-envelope protein \textit{SYNE1} (nesprin-1), the skeletal-muscle myosin heavy chain \textit{MYH4}, the Rho-associated coiled-coil kinase \textit{ROCK1}, the neural extracellular-matrix glycoprotein \textit{TNR} (tenascin-R), the epidermal barrier protein \textit{FLG} (filaggrin), and, with strong negative-direction effects, \textit{RBM39} and \textit{BRWD3}. Each reaches $\text{FDR} < 0.05$ against a volume- or organ-ratio-based imaging feature.

In HCC and BC (Fig.~\ref{fig:discovery}c,~d), no non-OncoKB gene crossed FDR; however, the top-ranked candidates by uncorrected $p$ include genes with exceptionally long coding sequences (\textit{OBSCN}, \textit{APOB}, \textit{PRUNE2} and \textit{RYR2} in liver; \textit{MACF1}, \textit{CACNA1C}, \textit{NEB} and \textit{CSMD3} in breast), several of which are recognised false-positive risks in mutation-burden analyses because their large genomic footprint inflates observed mutation counts relative to a uniform background rate \cite{nrg3539,mutational_landscape_drivers_2023}. Indeed, standard driver-detection pipelines explicitly exclude such long genes (including \textit{CSMD3} and \textit{LRP1B}) from significance testing to avoid this confound \cite{mutational_landscape_drivers_2023}. These nominally-significant lists are reported for completeness but are not claimed as discoveries.

Across all genes from all three cohorts, comparing the eight per-gene Evo~2 severity metrics reveals high concordance (Fig.~\ref{fig:discovery}f): the four $|\Delta\mathrm{LL}|$ magnitude summaries yield near-identical absolute partial-Spearman correlations within each gene, indicating that the observed gene--imaging associations are robust to the choice of severity aggregation. The raw mutation count ($N_{\mathrm{mut}}$) is the most frequent exception, consistent with the TMB adjustment removing much of its shared signal with imaging features. Together, these results demonstrate that the cross-modal sweep identifies novel gene--imaging associations, predominantly in cytoskeletal, ciliary and transporter biology, that recurrence statistics alone would miss.

\subsection{Pathway distribution of cRCC discoveries}

Mapping the 65 cRCC FDR-significant hits onto eleven curated KEGG functional groups (Fig.~\ref{fig:discovery}b) reveals an asymmetric distribution. The adhesion/cytoskeleton group accumulates the largest novel-gene contribution (5 of 6 hits novel, including \textit{ROCK1} and \textit{TNR}), followed by PI3K-Akt/mTOR (4 hits, evenly split between OncoKB-known and novel) and MAPK/Ras/RTK (5 hits, 3 OncoKB-known $+$ 2 novel). Pathways central to canonical cRCC biology, specifically HIF-1/VEGF, cell cycle/apoptosis/p53, and DNA-damage repair, were populated entirely by OncoKB-known genes, indicating that the novel-gene fraction is concentrated in cytoskeletal, calcium and signalling biology rather than in the classical tumor-suppressor pathways. Fifty of the 65 hits (13 OncoKB-known and 37 novel) fall outside the eleven curated groups, indicating a substantial off-pathway component to the discovery. In short, the canonical cRCC pathways are filled by known cancer genes, while the novel discoveries cluster in cytoskeletal biology and in the 50/65 hits that fall outside any of the eleven curated groups.

\subsection{Established disease relevance of novel cRCC genes}

Although the 46 novel hits are by definition absent from curated cancer-gene panels, they are not absent from human disease genetics. Cross-referencing the top-10 novel cRCC genes (the same ten shown in Fig.~\ref{fig:discovery}a) against both NCBI ClinVar \cite{landrum_clinvar_2018} and OMIM (Fig.~\ref{fig:discovery}e) shows that five are well-established Mendelian-disease genes with hundreds of pathogenic or likely-pathogenic (P/LP) ClinVar variants and multiple OMIM phenotypes: \textit{ALMS1} (1{,}048 P/LP alleles; Alstr\"{o}m syndrome, an autosomal-recessive ciliopathy \cite{rare_disease_77k_genomes_2023}), \textit{SYNE1} (398; cerebellar ataxia and Emery--Dreifuss muscular dystrophy), \textit{FLG} (247; atopic dermatitis), \textit{BRWD3} (69; X-linked intellectual disability), and \textit{TNR} (15; neurodevelopmental disorder). The remaining five (\textit{ABCA13}, \textit{MAST4}, \textit{MYH4}, \textit{ROCK1} and \textit{RBM39}) have no ClinVar or OMIM entries and are genuinely uncharacterised in clinical genetics. The over-representation of ciliopathy and cytoskeletal-organisation genes among the novel cRCC hits is consistent with the over-representation of adhesion/cytoskeleton and PI3K-Akt biology already visible in Fig.~\ref{fig:discovery}b, and positions these genes as candidate links between Mendelian ciliopathy/cytoskeletal disease biology and somatic kidney-cancer phenotype. Together, these results show that several of the novel cRCC genes are already known to cause inherited structural-biology disorders, suggesting that the same gene programmes may shape tumor morphology when disrupted somatically.

\section{Discussion}

By pairing Evo~2 severity scores with routine clinical imaging, our framework recovered established renal-cancer drivers in TCGA-cRCC at FDR-corrected significance and identified 46 genes absent from curated cancer-gene panels. Previous radiogenomic studies tested only pre-selected oncogenes \cite{0008_5472_can_17_0122v2} or small cohorts without validation \cite{radiogwas_pancreatic_2024}, and both depended on mutation recurrence, which cannot evaluate infrequently mutated genes \cite{somatic_mutation_rates_natbiotech_2022}. Because Evo~2 assigns a zero-shot log-likelihood drop to every variant based on evolutionary constraint \cite{s41586_026_10176_5_3,v1_covered_88832dba_a70d_41a6_b6bc_435eb56fee87}, the sweep is hypothesis-free and genome-wide in a sense that recurrence-dependent designs cannot achieve.

This reframes radiogenomics from confirmation to discovery. Earlier designs began with a list of cancer genes and asked which of them imprint on imaging; ours asks the inverse, of every mutated gene in the cohort, without pre-selecting any. The natural comparison is therefore with recurrence-based driver discovery rather than with prior radiogenomic work: recurrence statistics will keep finding the high-frequency drivers, but they cannot reach the long tail of low-recurrence genes that a per-variant severity score can evaluate directly.

Among the novel cRCC hits, ciliopathy and cytoskeletal genes are over-represented. \textit{ALMS1} (Alstr\"{o}m syndrome) \cite{rare_disease_77k_genomes_2023}, \textit{SYNE1} (nesprin-1; cerebellar ataxia), and \textit{ROCK1} (Rho-kinase; actin remodelling) all correlate with tumor volume, consistent with the role of primary cilia and the cytoskeleton in renal epithelial homeostasis. Their concentration in adhesion/cytoskeleton pathways, rather than in classical tumor-suppressor biology populated by known drivers, suggests the framework detects biology outside the established cancer-driver landscape. Perturbation experiments or (spatial) transcriptomics will be needed to clarify whether these associations are causal.

More broadly, the framework presented here opens a route toward systematic functional annotation of the thousands of somatically mutated genes that remain outside established cancer-gene panels. The human genome encodes roughly 20{,}000 protein-coding genes, and a substantial fraction still lacks reliable functional characterisation; research activity continues to concentrate on well-studied targets, leaving the ``unknome'' to shrink only slowly \cite{unknome_plos_biology_2023}. By replacing recurrence statistics with a foundation-model severity score, our approach removes the requirement for large numbers of carriers per gene and can, in principle, interrogate every mutated gene in a cohort. Because clinical imaging is acquired as part of routine care, extending the sweep to additional TCGA cancer types or to larger, uniformly imaged registries would require no new experimental data; the cost lies primarily in Evo~2 inference over the mutation catalogue. Such a scaled-up analysis could chart gene--phenotype associations across dozens of malignancies in a single pass, flagging candidates for targeted perturbation experiments or spatial transcriptomics follow-up. At the same time, ongoing improvements in genomic language models, including longer context windows and training on more diverse genomes \cite{s41586_026_10176_5_3,alphagenome_nature_2025}, are likely to sharpen the severity signal, while richer imaging features from volumetric deep-learning segmentations will capture phenotypic dimensions beyond volume alone. Combining these advances may transform the present proof-of-concept into a general-purpose discovery engine for previously uninvestigated gene--imaging links across cancer.

Nothing in the pipeline is intrinsically about imaging. The same Evo~2 score paired with TMB-adjusted partial correlation could be applied to digital pathology features, laboratory values, electronic health record (EHR)-derived phenotypes, or survival endpoints, given only a quantitative per-patient measurement to correlate against. Imaging is the most attractive starting point because it is collected at scale and at no marginal cost, but the same scaffold should generalise to any phenotype the genome plausibly shapes.

Several limitations apply. Only cRCC ($n = 162$) reached FDR significance; the smaller HCC and BC cohorts lacked power, and their nominally significant lists are affected by long-gene artefacts \cite{nrg3539,mutational_landscape_drivers_2023}. The Evo~2 log-likelihood drop reflects evolutionary constraint rather than oncogenic function; correlation with imaging does not establish causation. The analysis is restricted to exome data, missing non-coding variants that Evo~2 can score \cite{s41586_026_10176_5_3,alphagenome_nature_2025}, and heterogeneous TCGA imaging protocols introduce noise \cite{0008_5472_can_17_0122v2,radiogwas_pancreatic_2024}. Replication in larger, uniformly imaged cohorts \cite{dietlein_10478_cancer_genomes_natgen_2024} and extension to whole-genome sequencing will be essential next steps.

\section{Methods}

Our analysis proceeds in four stages (Fig.~\ref{fig:workflow}). First, we assemble three TCGA cohorts with paired somatic mutation calls and diagnostic imaging. Second, every somatic variant is scored with the Evo~2 genomic foundation model and the resulting per-variant severity values are aggregated into gene-level summary metrics. Third, radiomic features describing tumor size, shape, and attenuation are extracted from automated segmentations of the paired images. Finally, partial Spearman correlations, adjusted for tumor mutation burden, link each gene-level severity metric to each imaging feature, and Benjamini--Hochberg correction controls the false discovery rate across all tests. The subsections below detail each step.

\subsection{Cohorts and paired data}

The Cancer Genome Atlas (TCGA) is a public multi-omics resource that provides genomic profiling (whole-exome and whole-genome sequencing, RNA-seq, methylation arrays) together with clinical data for thousands of cancer patients across more than 30 tumor types \cite{pan_cancer_whole_genomes_2020}; matched diagnostic imaging for a subset of these patients is available through The Cancer Imaging Archive. We analysed three TCGA cohorts in which patients had both somatic mutation calls and tumor imaging available: TCGA-cRCC (clear cell renal cell carcinoma; $n = 162$ with paired data), TCGA-HCC (hepatocellular carcinoma; $n = 57$) and TCGA-BC (breast cancer; $n = 121$). Somatic mutation calls were obtained as Mutation Annotation Format (MAF) files (whole-exome sequencing with Ensembl Variant Effect Predictor annotation \cite{EMS140856}), all aligned to the GRCh38 (hg38) reference genome assembly. We verified that every MAF file used the same genome build so that genomic coordinates were consistent across cohorts and compatible with the reference sequence context required for Evo~2 scoring. The corresponding reference and mutant allele contexts were extracted from the GRCh38 assembly for each variant position. Tumor segmentations were obtained with cohort-specific pre-trained models: the KiTS23 \cite{heller_kits_2023} nnU-Net \cite{isensee_nnunet_2021} v1 5-fold ensemble for TCGA-cRCC; TotalSegmentator \cite{wasserthal_totalsegmentator_2023} for TCGA-HCC contrast-enhanced CT, with quality filtering excluding scans whose predicted total liver volume was $< 500$~mL; and the MAMA-MIA \cite{garrucho_mamamia_2024} nnU-Net v2 model applied to the post-contrast phase of TCGA-BC dynamic contrast-enhanced magnetic resonance imaging (DCE-MRI) series.

\subsection{Per-mutation Evo~2 severity scoring}

For each somatic single-nucleotide variant, we computed the Evo~2 \cite{s41586_026_10176_5_3} likelihood drop $|\Delta\mathrm{LL}|$ as the absolute difference in the model's log-likelihood of the reference and mutant alleles within their $\pm 4{,}096$~bp genomic context. Larger $|\Delta\mathrm{LL}|$ denotes a sequence less plausible under the genomic-language-model prior, used here as a foundation-model proxy for functional disruption of the encoded gene element. Due to the nature of targeted exome sequencing, which inherently limits detection to coding regions and may exclude intronic or structural variants, all somatic variants were analysed at the single-allele level; consequently, we excluded zygosity information from our computations.

\subsection{Per-gene aggregation}

For each gene with at least five mutation carriers in a cohort, eight per-gene severity summaries were computed across that gene's mutations within each carrier: maximum $|\Delta\mathrm{LL}|$, mean $|\Delta\mathrm{LL}|$, sum $|\Delta\mathrm{LL}|$, standard deviation of $|\Delta\mathrm{LL}|$, signed minimum $\Delta\mathrm{LL}$, signed maximum $\Delta\mathrm{LL}$, the signed $\Delta\mathrm{LL}$ of the worst-$|\Delta\mathrm{LL}|$ mutation in the gene (top-1 $\Delta\mathrm{LL}$), and the mutation count ($n_{\mathrm{mutations}}$). Non-carriers received zero for severity summaries and zero for $n_{\mathrm{mutations}}$.

\subsection{Tumor radiomic features}

Per-patient features extracted from the tumor segmentations included tumor volume ($V_{\mathrm{tumor}}$), tumor-to-organ volume ratio, and intra-mask intensity heterogeneity (coefficient of variation of Hounsfield Unit (HU) / MR signal), with cohort-specific additions: necrotic fraction (cRCC, HCC) and intra-mask intensity mean and standard deviation (BC tumor; HCC liver). For TCGA-BC, the ``organ'' denominator is the field-of-view (FOV) non-zero MR volume rather than a dedicated whole-breast segmentation, and the resulting tumor-to-breast ratio should be read as an approximate tumor-to-FOV ratio.

\subsection{Cross-modal correlation}

For each combination of gene, severity metric and imaging feature, we computed a TMB-adjusted partial Spearman correlation: the per-gene metric and the imaging feature were each residualized by ordinary least squares on the patient's total mutation count before the rank correlation, removing the cohort-wide confounding of mutation burden with tumor burden. We restricted to combinations with at least 30 patients in the overlapping support and at least five mutation carriers having imaging.

\subsection{Multiple-testing correction}

Within each cohort, $p$-values from all per-gene metric--imaging feature tests (2{,}984 in cRCC, 2{,}220 in BC, 1{,}638 in HCC) were adjusted for multiple comparisons using the Benjamini--Hochberg procedure \cite{benjamini_hochberg_fdr_1995}, which controls the false discovery rate (FDR), defined as the expected proportion of rejected null hypotheses that are false positives. Each test receives an adjusted $p$-value, termed a q-value: a q-value of 0.05 for a given gene--imaging combination means that, among all combinations with equal or smaller q-values, no more than 5\% are expected to be false discoveries. A combination was declared significant at $q < 0.05$. The q-values used to declare significance came from the full 2{,}984-test family, not from the 65-element best-per-gene subset.

\subsection{External annotations}

To contextualise the FDR-significant gene set, we cross-referenced each gene against three external resources. A gene was labelled OncoKB-known if it appeared in any of the seven panels distributed in the OncoKB cancer-gene compendium \cite{chakravarty_oncokb_2017} (approximately 1{,}236 genes in total), comprising the OncoKB-annotated list, MSK-IMPACT, MSK-HEME, FoundationOne, FoundationOne-HEME, the Vogelstein driver list, and the COSMIC Cancer Gene Census v99. For pathway-level interpretation, we defined eleven curated high-level functional groups (calcium signalling; MAPK/Ras/RTK; PI3K-Akt/mTOR; Wnt/Notch/Hedgehog; hormone; HIF-1/VEGF; TGF-$\beta$; DNA damage/repair; cell cycle/apoptosis/p53; ubiquitin/proteasome; adhesion/cytoskeleton), each constructed as the union of member genes from one or more KEGG pathways \cite{kanehisa_kegg_2000} resolved via the KEGG REST API. Finally, per-gene counts of pathogenic and likely-pathogenic alleles were obtained from the NCBI ClinVar database \cite{landrum_clinvar_2018} (April 2026 release), and the primary curated disease association per gene was defined as the shortest distinct disease name linked to that gene across both direct and related gene entries.

\subsection{Statistics and reproducibility}

All statistical tests are two-sided unless stated otherwise. The Benjamini--Hochberg procedure was used to control the false discovery rate at $q < 0.05$ within each cohort. Fisher's exact test (one-sided) was used for OncoKB enrichment. No data were excluded from the analysis except as described above (quality filtering of HCC scans with predicted liver volume $< 500$~mL). No randomisation or blinding was applicable to this retrospective computational study.

\subsection{Reporting summary}

Further information on research design is available in the Nature Portfolio Reporting Summary linked to this article.

\section*{Data availability}

Somatic mutation calls (MAF files) and clinical metadata are available from the Genomic Data Commons (\url{https://portal.gdc.cancer.gov}) for TCGA-cRCC, TCGA-HCC and TCGA-BC. Diagnostic imaging is available from The Cancer Imaging Archive (\url{https://www.cancerimagingarchive.net}).

\section*{Code availability}

Code for Evo~2 scoring, radiomic feature extraction and the cross-modal correlation analysis is available at \url{https://github.com/TruhnLab/OncoGenesRadiologyDiscovery}.

\section*{Author contributions}

F.H. conceived the study, developed the analysis pipeline, performed all experiments and wrote the manuscript. J.K. contributed to data analysis and interpretation of results, and reviewed the manuscript. S.H., P.W., C.K., I.K., J.N.K. and S.N. reviewed the manuscript and provided critical feedback. D.T. supervised the project and revised the manuscript.

\section*{Competing interests}

D.T. holds shares in StratifAI and Synagen. He has received honoraria from Bayer, AstraZeneca, Philips, Roche, Pfizer, and Gilead. J.N.K. declares ongoing consulting services for AstraZeneca, Panakeia, and Bioptimus. He holds shares in StratifAI, Synagen, and Spira Labs, has received an institutional research grant from GSK and AstraZeneca, as well as honoraria from AstraZeneca, Bayer, Daiichi Sankyo, Eisai, Janssen, Merck, MSD, BMS, Roche, Pfizer, and Fresenius. All other authors declare no conflicts of interest.

\section*{Materials \& Correspondence}

Correspondence and requests for materials should be addressed to F.H. (fhauke@ukaachen.de).


\begin{thebibliography}{99}

\bibitem{pan_cancer_whole_genomes_2020}
The ICGC/TCGA Pan-Cancer Analysis of Whole Genomes Consortium.
Pan-cancer analysis of whole genomes.
\textit{Nature} \textbf{578}, 82--93 (2020).

\bibitem{nrg3539}
Watson, I.~R., Takahashi, K., Futreal, P.~A. \& Chin, L.
Emerging patterns of somatic mutations in cancer.
\textit{Nat.\ Rev.\ Genet.} \textbf{14}, 703--718 (2013).

\bibitem{mutational_landscape_drivers_2023}
Sinkala, M.
Mutational landscape of cancer-driver genes across human cancers.
\textit{Sci.\ Rep.} \textbf{13}, 12742 (2023).

\bibitem{dietlein_10478_cancer_genomes_natgen_2024}
Kinnersley, B., Sud, A., Everall, A. et al.
Analysis of 10,478 cancer genomes identifies candidate driver genes and opportunities for precision oncology.
\textit{Nat.\ Genet.} \textbf{56}, 1868--1877 (2024).

\bibitem{somatic_mutation_rates_natbiotech_2022}
Sherman, M.~A. et al.
Genome-wide mapping of somatic mutation rates uncovers drivers of cancer.
\textit{Nat.\ Biotechnol.} \textbf{40}, 1634--1643 (2022).

\bibitem{unknome_plos_biology_2023}
Rocha, J.~J. et al.
Functional unknomics: systematic screening of conserved genes of unknown function.
\textit{PLoS Biol.} \textbf{21}, e3002222 (2023).

\bibitem{rare_disease_77k_genomes_2023}
Greene, D. et al.
Genetic association analysis of 77,539 genomes reveals rare disease etiologies.
\textit{Nat.\ Med.} \textbf{29}, 679--688 (2023).

\bibitem{radiomics_tumor_phenotype_natcomm_2014}
Aerts, H.~J.~W.~L. et al.
Decoding tumour phenotype by noninvasive imaging using a quantitative radiomics approach.
\textit{Nat.\ Commun.} \textbf{5}, 4006 (2014).

\bibitem{0008_5472_can_17_0122v2}
Rios Velazquez, E. et al.
Somatic mutations drive distinct imaging phenotypes in lung cancer.
\textit{Cancer Res.} \textbf{77}, 3922--3930 (2017).

\bibitem{radiogwas_pancreatic_2024}
Zheng, D. et al.
radioGWAS links radiome to genome to discover driver genes with somatic mutations for heterogeneous tumor image phenotype in pancreatic cancer.
\textit{Sci.\ Rep.} \textbf{14}, 12316 (2024).

\bibitem{nucleotide_transformer_2024}
Dalla-Torre, H., Gonzalez, L., Mendoza-Revilla, J. et al.
Nucleotide Transformer: building and evaluating robust foundation models for human genomics.
\textit{Nat.\ Methods} \textbf{22}, 287--297 (2025).

\bibitem{alphagenome_nature_2025}
Avsec, \v{Z}., Latysheva, N., Cheng, J. et al.
Advancing regulatory variant effect prediction with AlphaGenome.
\textit{Nature} \textbf{649}, 1206--1218 (2026).

\bibitem{s41586_026_10176_5_3}
Brixi, G. et al.
Genome modelling and design across all domains of life with Evo~2.
\textit{Nature} \textbf{652}, 1349--1361 (2026).

\bibitem{v1_covered_88832dba_a70d_41a6_b6bc_435eb56fee87}
Ohno-Machado, L., Zhu, R., Zhou, X. et al.
Advancing human population genomics with DNA foundation models.
Preprint at \textit{Research Square} (2025).

\bibitem{isensee_nnunet_2021}
Isensee, F., Jaeger, P.~F., Kohl, S.~A.~A., Petersen, J. \& Maier-Hein, K.~H.
nnU-Net: a self-configuring method for deep learning-based biomedical image segmentation.
\textit{Nat.\ Methods} \textbf{18}, 203--211 (2021).

\bibitem{wasserthal_totalsegmentator_2023}
Wasserthal, J. et al.
TotalSegmentator: robust segmentation of 104 anatomic structures in CT images.
\textit{Radiol.\ Artif.\ Intell.} \textbf{5}, e230024 (2023).

\bibitem{heller_kits_2023}
Heller, N. et al.
The KiTS21 challenge: automatic segmentation of kidneys, renal tumors, and renal cysts in corticomedullary-phase CT.
Preprint at \textit{arXiv} 2307.01984 (2023).

\bibitem{garrucho_mamamia_2024}
Garrucho, L. et al.
A large-scale multicenter breast cancer DCE-MRI benchmark dataset with expert segmentations.
\textit{Sci.\ Data} \textbf{12}, 453 (2025).

\bibitem{chakravarty_oncokb_2017}
Chakravarty, D. et al.
OncoKB: a precision oncology knowledge base.
\textit{JCO Precis.\ Oncol.} \textbf{1}, PO.17.00011 (2017).

\bibitem{tcga_kirc_nature_2013}
Cancer Genome Atlas Research Network.
Comprehensive molecular characterization of clear cell renal cell carcinoma.
\textit{Nature} \textbf{499}, 43--49 (2013).

\bibitem{landrum_clinvar_2018}
Landrum, M.~J. et al.
ClinVar: improving access to variant interpretations and supporting evidence.
\textit{Nucleic Acids Res.} \textbf{46}, D1062--D1067 (2018).

\bibitem{kanehisa_kegg_2000}
Kanehisa, M. \& Goto, S.
KEGG: Kyoto Encyclopedia of Genes and Genomes.
\textit{Nucleic Acids Res.} \textbf{28}, 27--30 (2000).

\bibitem{benjamini_hochberg_fdr_1995}
Benjamini, Y. \& Hochberg, Y.
Controlling the false discovery rate: a practical and powerful approach to multiple testing.
\textit{J.~R.~Stat.\ Soc.\ Ser.~B} \textbf{57}, 289--300 (1995).

\bibitem{EMS140856}
Hunt, S.~E., Moore, B., Amode, R.~M. et al.
Annotating and prioritizing genomic variants using the Ensembl Variant Effect Predictor---a tutorial.
\textit{Hum.\ Mutat.} \textbf{43}, 986--997 (2022).

\end{thebibliography}
\end{document}